\documentclass[fleqn,usenatbib]{mnras}

\usepackage[T1]{fontenc}
\usepackage{ae,aecompl}

\usepackage{graphicx}	
\usepackage{amsmath}	
\usepackage{amssymb}	
\usepackage{slashed}    
\DeclareMathOperator*{\argminT}{argmin}
\DeclareMathOperator*{\argmaxT}{argmax}

\usepackage{mathtools}
\DeclarePairedDelimiterX{\norm}[1]{\lVert}{\rVert}{#1}
\makeatletter
\newcommand*{\rom}[1]{\expandafter\@slowromancap\romannumeral #1@}
\makeatother
\usepackage{wrapfig}
\usepackage{bm}
\usepackage{multirow}
\usepackage[utf8]{inputenc}
\usepackage{soul}

\defcitealias{[32]}{J14}
\defcitealias{[36]}{C12}

\usepackage{smartdiagram}
\usepackage{tikz}
\usetikzlibrary{shapes,arrows}
\tikzstyle{block} = [rectangle, draw, fill=blue!10,
    text width=12em, text centered, rounded corners, minimum height=3em]
\tikzstyle{block2} = [rectangle, draw, fill=purple!10,
    text width=12em, text centered, rounded corners, minimum height=3em]
\tikzstyle{line} = [draw, -latex']
\tikzstyle{decision} = [diamond, draw, fill=purple!10,
    text width=4.5em, text badly centered, node distance=2.2cm, inner sep=0pt]

\newcommand{\emdash}{\nobreak--\nobreak\hskip3pt}

\usepackage{algorithm}
\usepackage{setspace}
\usepackage[noend]{algpseudocode}
\makeatletter
\def\BState{\State\hskip-\ALG@thistlm}
\makeatother

\usepackage{etoolbox}

\usepackage{newtxtext,newtxmath}

\title[Mass-mapping: hypothesis testing]{Sparse Bayesian mass-mapping with uncertainties: hypothesis testing of structure}

\author[Price et al.]{
M.~A.~Price$^{1}$\thanks{E-mail: m.price.17@ucl.ac.uk}, J.~D.~McEwen$^{1}$, X.~Cai$^{1}$, T.~D.~Kitching$^{1}$, C.~G.~R.~Wallis$^{1}$ \newauthor
\normalsize(for the LSST Dark Energy Science Collaboration)
\\
$^{1}$Mullard Space Science Laboratory, University College London, RH5 6NT, UK.\\
}

\date{Accepted XXX. Received YYY; in original form ZZZ}

\pubyear{2018}

\begin{document}
\label{firstpage}
\pagerange{\pageref{firstpage}--\pageref{lastpage}}
\maketitle

\begin{abstract}
    A crucial aspect of mass-mapping, via weak lensing, is quantification of
    the uncertainty introduced during the reconstruction process. Properly
    accounting for these errors has been largely ignored to date. We present
    a new method to reconstruct \textit{maximum a posteriori} (MAP) convergence
    maps by formulating an unconstrained Bayesian inference problem with
    Laplace-type $l_1$-norm sparsity-promoting priors, which we solve via
    convex optimization. Approaching mass-mapping in this manner allows us to
    exploit recent developments in probability concentration theory to infer
    theoretically conservative uncertainties for our MAP reconstructions,
    without relying on assumptions of Gaussianity.  For the first time these
    methods allow us to perform hypothesis testing of structure, from which
    it is possible to distinguish between physical objects and artifacts of
    the reconstruction. Here we present this new formalism, demonstrate the
    method on simulations, before applying the developed formalism
    to two observational datasets of the Abel-520 cluster. Initial
            reconstructions of the Abel-520 catalogs reported the detection of an
            anomalous `dark core' -- an over dense region with no optical counterpart --
            which was taken to be evidence for self-interacting dark-matter. In our
    Bayesian framework it is found that neither Abel-520 dataset can conclusively
    determine the physicality of such dark cores at $99\%$ confidence. However, in both cases the recovered MAP estimators
    are consistent with both sets of data.
\end{abstract}



\section{Introduction}

Gravitational lensing is an astrophysical phenomenon, that can be observed
on galactic to cosmic spatial scales, through which distant images are
distorted by the intervening mass density field.  Due to its nature, lensing
is sensitive to the total mass distribution (both visible and invisible)
along a line of sight \citep{[1],[2],[23],[24]}. Therefore, as the majority
of massive structures in the universe are predominantly dark matter, lensing
provides a novel way to probe the nature of dark matter itself.
Weak gravitational lensing (WL) is a regime in which one makes the approximation
that lensed sources have (at no time) come radially closer than an Einstein
radius to the intervening mass concentrations \emdash which ensures that
sources are not multiply imaged. The effect of weak lensing on distant source
galaxies is two-fold: the galaxy size is magnified by a convergence field
$\kappa$; and the galaxy ellipticity (third-flattening) is perturbed from an
underlying intrinsic value by a shearing field $\gamma$.
\par
Due to the mass-sheet degeneracy the weak lensing convergence field is not directly observable.
In the weak lensing regime, the shearing field does not suffer such
degeneracies and can accurately be modelled from observed ellipticities.
Therefore, observations of $\gamma$ are typically inverted to recover
estimators of $\kappa$.  Such estimators are colloquially
named \textit{dark matter mass-maps}, and constitute one of the principle
observables for cosmology \citep{[25]}.
Standard cosmological protocol is to extract weak lensing information in
the form of second order statistics \citep{[26],[49],kilbinger2015} which
are then compared to theory. In this approach mass-maps are not required.
However, as two-point global statistics are by definition sensitive only to
Gaussian contributions, and weak lensing is inherently non-Gaussian, it is
informative to consider higher-order statistics \citep{[27], [28]}. Many
higher-order statistical techniques can be performed directly on mass-maps
($\kappa$-fields), which motivates investigation into alternate mass-map
reconstruction methodologies.
\par
Reconstructing mass-maps from shear observations requires solving an ill-posed
(often seriously) inverse problem. Many approaches to solving this lensing
inverse problem have been developed \citep[\emph{e.g.}][]{[29],[5],[6],[3],[15],[22]}, 
with the industry standard being Kaiser-Squires \citep[KS,][]{[5]}.
Although these approaches often produce reliable convergence estimators,
they lack principled statistical approaches to uncertainty quantification
and often assume Gaussianty during the reconstruction
process, or post-process by Gaussian smoothing, which is sub-optimal when one wishes to analyze small-scale non-Gaussian
structure.
\par
Most methods refrain from quantifying uncertainties in reconstructions, but
those that do often do so by assuming Gaussian priors and adopting Markov-chain
Monte-Carlo (MCMC) techniques \citep{[30],[26],[43]}.The computational cost of
MCMC approaches is large. Recent developments in probability concentration theory
have led to advancements in fast approximate uncertainty quantification
techniques \citep{[10],[11],[12]}.
\par
In this article we present a new mass-mapping formalism. We formulate the
lensing inverse problem as a sparse hierarchical Bayesian inference problem
from which we derive an unconstrained convex optimization problem. We solve
this optimization problem in the analysis setting, with a wavelet-based,
sparsity-promoting, $l_1$-norm prior \emdash similar priors have been shown
to be effective in the weak lensing setting \citep{[15],[6],[9],[46]}. Formulating
the problem in this way allows us, for the first time, to recover \textit{maximum
    a posteriori} (MAP) estimators, from which we can exploit analytic methods
\citep{[10],[12]} to recover approximate highest posterior density (HPD)
credible regions, and perform hypothesis testing of structure in a variety of
ways. We apply our algorithm to a range of catalogs drawn from N-body
        simulations \emdash Bolshoi  cluster catalogues \citep{[20]} \emdash and the
debated A520 cluster catalogs \citep{[36], [32]}. We then demonstrate the
aforementioned uncertainty quantification techniques on our MAP reconstructions
from these catalogs.
\par
The structure of this article is as follows. In section \ref{Background} we
provide a brief overview of the weak lensing paradigm and motivate a sparsity-based
approach. In section \ref{Sparsity} we provide the details of our algorithm,
as well as some updates to super-resolution image recovery. In section \ref{Bayesian
    Uncertainties} we present the uncertainty quantification techniques, both
mathematically and mechanistically. In sections \ref{Simulation Results} and
\ref{A520} we apply both our reconstruction algorithm and the uncertainty
quantification techniques to the aforementioned datasets and analyze the
results. Finally, in section \ref{Conclusions} we draw conclusions from this
work and propose future avenues of research.
\par
Section \ref{Sparsity} relies on a moderate level of understanding in the
fields of proximal calculus and compressed sensing, and section \ref{Bayesian
    Uncertainties} relies on a general understanding of Bayesian inference. As such,
for the reader solely interested in practical application of these techniques
we recommend sections \ref{Simulation Results} onwards.

\section{Weak Gravitational Lensing} \label{Background}
The following section presents a brief review of the mathematical background
relevant to the weak lensing formalism, though a deeper description can be
found in popular review articles \citep{[1],[2]}.

\subsection{Weak lensing regime}
Gravitational lensing refers to the deflection of distant photons as they
propagate from their origin to us, the observer. This deflection is caused by
local Newtonian potentials which are, in turn, sourced by the total local
matter over or under density. As such, weak lensing is sensitive to both the
visible and invisible matter distribution \emdash making it an ideal probe of
dark matter in the Universe.
\par
The weak gravitational lensing regime is satisfied when propagating photons
(from a distant source) have an angular position on the source plane $\beta$
(relative to the line-of-sight from observer through the lensing mass) greater
than the Einstein radius $\theta_E$ of the intervening mass. This assertion
ensures that the solution of the first order lens equation is singular:
\begin{equation}
    \beta = \theta - \theta_E^2 \frac{\theta}{|\theta|^2}.
\end{equation}
Where the Einstein radius is defined to be:
\begin{equation}
    \theta_E = \sqrt{\frac{4GM}{c^2}\frac{f_K(r - r^{\prime})}{f_K(r)f_K(r^{\prime})}},
\end{equation}
where $f_K$ is the angular diameter distance in a cosmology with curvature
$K$, $c$ is the speed of light in a vacuum, $G$ is the gravitational constant
and $M$ is the lensing mass. Perhaps more generally the weak lensing regime
can be defined as convergence fields for which $\kappa \ll 1$ -- ensuring
that the shear signal remains linear.
\par
Due to the sparse nature of the distribution of galaxies across the sky, most
sources are (to a good approximation) within the weak gravitational lensing
regime. The weak gravitational lensing effect is best expressed in terms of
a lensing potential $\phi$, defined to be the integral of the Newtonian
potential $\Phi$ along a given line of sight:
\begin{equation}
    \phi(r,\omega) = \frac{2}{c^2} \int_0^r dr^{\prime} \frac{f_K(r - r^{\prime})}{f_K(r) f_K(r^{\prime})} \Phi(r^{\prime}, \omega),
\end{equation}
where $r$ and $r^{\prime}$ are co-moving distances, and $\omega = (\theta,\psi)$
are angular spherical co-ordinates. The local Newtonian potential must satisfy
the Poisson equation and as such is related to the matter over-density field:
\begin{equation}
    \nabla^2 \Phi(r,\omega) = \frac{3 \Omega_M H_0^2}{2a(r)} \delta(r,\omega),
\end{equation}
where $\Omega_M$ is the matter density parameter, $H_0$ is the current Hubble
constant, $a(r)$ is the scale factor, and $\delta$ is the fractional over-density.
\par
To first order, there are two primary ways in which light from distant sources
is distorted by this lensing potential. Images are magnified by a spin-0
convergence field $\kappa$ and sheared by a spin-2 shear field $\gamma$. These
quantities can be shown \citep{[1]} to be related to  the lensing potential by:
\begin{align}
     & \kappa(r,\omega) = \frac{1}{4}(\eth \bar{\eth} + \bar{\eth} \eth) \; \phi(r,\omega), \label{eq:kappatophi} \\
     & \gamma(r,\omega) = \frac{1}{2} \eth \eth \; \phi(r,\omega),\label{eq:gammatophi}
\end{align}
where $\eth$ and $\bar{\eth}$ are the spin $s$ raising and lowering operators
respectively and are in general defined to be,
\begin{align}
    \eth \equiv -\sin^s\theta \Big ( \frac{\partial}{\partial \theta} + \frac{i \partial}{\sin\theta \partial \psi} \Big ) \sin^{-s}\theta, \\
    \bar{\eth} \equiv -\sin^{-s}\theta \Big ( \frac{\partial}{\partial \theta} - \frac{i \partial}{\sin\theta \partial \psi} \Big ) \sin^{s}\theta.
\end{align}
Where we have omitted spin subscripts for clarity.

\subsection{Standard mass-mapping techniques}
Typically we wish to make inferences about the projected matter over-density
$\delta(r,\omega)$ which is most directly accessible by inverting the integral
equation \citep{[2]}
\begin{equation}
    \kappa(r, \omega) = \frac{3\Omega_M H_0^2}{2c^2} \int_0^{r} d r^{\prime}\frac{f_K(r^{\prime})f_K(r - r^{\prime})}{f_K(r)} \frac{\delta(f_K(r^{\prime})r^{\prime}, r^{\prime})}{a(r)}.
\end{equation}
\par
This poses a difficulty as the convergence $\kappa$ is only determined to the
degeneracy $\kappa \rightarrow \kappa^\prime = \eta \kappa + (1-\eta)$ and is therefore not
directly observable --- this degeneracy is often referred to as the \textit{mass-sheet degeneracy}.
However, as the intrinsic ellipticity distribution of galaxies has zero mean, if one averages
many galaxy ellipticities within a given pixel the true shear $\gamma$ can be
recovered -- which makes $\gamma$ an observable field. As such one typically collects
observations of $\gamma$ which are and subsequently used to construct estimators of $\kappa$.
\par
For small sky fractions we can approximate the field of view as a plane (though
this approximation degrades quickly with sky fraction; \citeauthor{[3]}
\citeyear{[3]}). In this planar approximation $\eth$ and $\bar{\eth}$ reduce to
\citep{[4]}:
\begin{equation}
    \eth \approx -(\partial_x + i \partial_y) \quad \mbox{and} \quad \bar{\eth} \approx -(\partial_x - i \partial_y).
\end{equation}
Combining equations (\ref{eq:kappatophi}) and (\ref{eq:gammatophi}) we find the
planar forward model in Fourier space:
\begin{equation} \label{eq:forwardmodel}
    \hat{\gamma}(k_x,k_y) = \bm{\mathsf{D}}_{k_x,k_y} \hat{\kappa}(k_x, k_y),
\end{equation}
with the mapping operator being,
\begin{equation} \label{eq:KSD}
    \bm{\mathsf{D}}_{k_x,k_y} = \frac{k_x^2-k_y^2+2ik_xk_y}{k_x^2+k_y^2}.
\end{equation}
Hereafter we drop the $k_x,k_y$ subscripts for clarity. It is informative to
note that this forward model is undefined at the origin ($k=\sqrt{k_x^2+k_y^2}=0$)
\emdash which corresponds to the mass-sheet degeneracy \citep{[1]}
\par
The most naive inversion of this forward model is Kaiser-squires (KS) inversion,
\begin{equation} \label{eq:Kaiser-Squires}
    \hat{\kappa}^{\text{KS}} = \bm{\mathsf{D}}^{-1} \hat{\gamma},
\end{equation}
which is direct inversion in Fourier space \citep{[5]}. KS inversion of the
forward model, given by equation (\ref{eq:forwardmodel}), performs adequately,
provided the space over which it is defined is complete, and the sky fraction
is small. However, masking and survey boundaries are inherent in typical weak
gravitational lensing surveys, leading to significant contamination of the KS
estimator. Often maps recovered with the KS estimator are convolved with a Gaussian
kernel to reduce the impact of these contaminations but this is sub-optimal. This
smooths away a large fraction of the small-scale non-Gaussian information, which
cosmologists are increasingly interested in extracting from weak gravitational
lensing surveys.

\section{Sparse MAP Estimators} \label{Sparsity}
Several alternate approaches for solving the inverse problem between
convergence $\kappa$ and shear $\gamma$ which do not assume or impose Gaussianity
have been proposed, some of which are based on the concept of wavelets and
sparsity \citep{[6], [7], [8], [9]}.
\par
We propose a mass-mapping algorithm that relies on sparsity in a given wavelet
dictionary. Moreover, we formulate the problem such that we can exploit recent
developments in the theory of probability concentration, which have been developed
further to produce novel uncertainty quantification techniques \citep{[10]}.
Crucially, this allows us to recover principled statistical uncertainties on our
MAP reconstructions \citep[as in][]{[11], [12]} as will be discussed in detail
in the following section.
\par
As mentioned previously, galaxies have an intrinsic ellipticity. To mitigate
the effect of intrinsic ellipticity we choose to project the ellipticity measurements
onto a grid and average. If we assume that galaxies have no preferential orientation
in the absence of lensing effects, then the average intrinsic ellipticity tends to
zero. This is a good approximation for the purposes of this paper, but weak
correlation between the intrinsic alignments of galaxies has been observed
\citep{[13],[45]}.

\subsection{Hierarchical Bayesian Framework}
Hierarchical Bayesian inference provides a rigorous mathematical framework through
which theoretically optimal solutions can be recovered. Moreover it allows one to
construct measures of the uncertainty on recovered point estimates.
\par
As is common for hierarchical Bayesian models, we begin from Bayes' theorem for the
posterior distribution,
\begin{equation} \label{eq:bayes}
    p(\kappa|\gamma) = \frac{p(\gamma|\kappa)p(\kappa)}{\int_{\mathbb{C}^N} p(\gamma|\kappa)p(\kappa)d\kappa},
\end{equation}
where $p(\gamma|\kappa)$ is the likelihood function representing data fidelity,
$N$ is the dimensionality of $\kappa$ and $p(\kappa)$ is a prior on the statistical
nature of $\kappa$. The denominator is called the \textit{Bayesian evidence} which
is constant and so can be dropped for our purposes. Typically the Bayesian evidence
is used for model comparison, which we will not be considering within the context of
this paper. Given Bayes' theorem, and the monotonicity of the logarithm function, we
can easily show that the maximum posterior solution is defined by,
\begin{equation} \label{eq:logposterior}
    \argmaxT_{\kappa} \lbrace p(\kappa|\gamma) \rbrace = \argminT_{\kappa} \lbrace -\log ( \; p(\kappa|\gamma) \;) \rbrace.
\end{equation}
This step is crucial, as it allows us to solve the more straightforward problem of
minimizing the log-posterior rather than maximizing the full posterior. Conveniently,
in most physical situations the operators associated with the log-posterior are
convex. Drawing from the field of convex optimization, the optimal solution for the
posterior can be recovered extremely quickly \emdash even in high dimensional
settings.

\subsection{Sparsity and Inverse problems}
Let $\gamma \in \mathbb{C}^M$ be the discretized complex shear field
extracted from an underlying discretized convergence field $\kappa \in \mathbb{C}^N$
by a \textit{measurement operator} $\bm{\Phi} \in \mathbb{C}^{M \times N}: \kappa
    \mapsto \gamma$. In the planar setting $\bm{\Phi}$ can be modeled by,
\begin{equation} \label{eq:measurement_operator}
    \bm{\Phi} = \bm{\mathsf{M}} \bm{\mathsf{F}}^{-1} \bm{\mathsf{D}} \bm{\mathsf{F}}.
\end{equation}
Here $\bm{\mathsf{F}}$ is the discrete fast Fourier transform (FFT),
$\bm{\mathsf{F}}^{-1}$ is the inverse discrete fast Fourier transform (IFFT),
$\bm{\mathsf{M}}$ is a standard masking operator, and $\bm{\mathsf{D}}$ is a
diagonal matrix applying the scaling of the forward model in Fourier space as
defined in equation (\ref{eq:KSD}). In the case of independent and identically
distributed \textit{i.i.d.} Gaussian noise, measurement of $\gamma$ will be
contaminated such that:
\begin{equation}
    \gamma = \bm{\Phi} \kappa + \mathcal{N}(0,\sigma_{i}^2),
\end{equation}
where $\mathcal{N}(0,\sigma_{i}^2) \in \mathbb{C}^M$ is additive i.i.d. Gaussian
noise of variance $\sigma_{i}^2$ for pixel $i$. Often in weak gravitational
lensing experiments the total number of binned measurements is less than the
number of pixels to be recovered, $M < N$, and the inverse problem becomes ill-posed.
\par
In such a setting the Bayesian likelihood function (data fidelity term) is given by the
product of Gaussian likelihoods defined on each pixel with pixel noise variance
$\sigma_i^2$, which is to say an overall multivariate Gaussian likelihood of known
covariance $\Sigma = \text{diag}(\sigma_1, \sigma_2,\dots,\sigma_M) \in \mathbb{R}^{M \times M}$.
Let $\Phi_i\kappa$ be the value of $\bm{\Phi}\kappa$ at pixel $i$, then the overall
likelihood is then defined as,
\begin{align}
    p(\gamma|\kappa) & \propto \prod_{i=0}^{M} \exp \Bigg(\frac{-( \Phi_i \kappa - \gamma_i )^2}{2\sigma_i^2} \Bigg)
    = \prod_{i=0}^{M} \exp \Bigg( \frac{-1}{2} \Big ( \bar{\Phi}_i \kappa - \bar{\gamma}_i \Big )^2 \Bigg), \nonumber                                                  \\
                     & = p(\gamma | \kappa) \propto \exp \Bigg( \frac{-\norm{ \bar{\bm{\Phi}} \kappa - \bar{\gamma} }_2^2}{2} \Bigg), \label{eq:covariance_likelihood}
\end{align}
where $\norm{ \cdot}_2$ is the $\ell_2$-norm and $\bar{\bm{\Phi}} = \Sigma^{-\frac{1}{2}}\bm{\Phi}$
is a composition of the measurement operator and an inverse covariance weighting.
Effectively this covariance weighting leads to measurements $\bar{\gamma} = \Sigma^{-\frac{1}{2}}\gamma$
which whiten the typically non-uniform noise variance in the observational data $\gamma$.
\par
This likelihood function allows one to map from the  number count of observations
per pixel to a corresponding noise variance (assuming an intrinsic ellipticity dispersion of $\sim 0.37$),
from which the noise (under and central limit theory argument of Gaussianity) may be correctly
incorporated into the reconstruction. In practice this requires only the number density of
observations per pixel, which is trivially inferred from raw observational data catalogues.
\par
To regularize this inverse problem, we then define a sparsity promoting Laplace-type prior:
\begin{equation}
    p(\kappa) \propto \exp \bigg(-\mu \norm{\bm{\Psi}^{\dag}\kappa}_1 \bigg),
\end{equation}
where $\bm{\Psi}$ is an appropriately selected wavelet dictionary, and $\mu \in
    \mathbb{R}_{+}$ is a regularization parameter \emdash effectively a weighting
between likelihood and prior. Note that one may choose any convex log-prior
within this formalism \textit{e.g.} an $\ell_2$-norm prior from which one
essentially recovers Weiner filtering \citep[see][for alternate iterative Weiner
    filtering approaches]{Seljak2003,Horowitz2018}. From equations (\ref{eq:bayes})
and (\ref{eq:logposterior}) the unconstrained optimization problem which minimizes
the log-posterior is,
\begin{equation} \label{eq:optimization-problem}
    \kappa^{\text{map}} = \argminT_{\kappa} \Bigg \lbrace \mu \norm{ \bm{\Psi}^{\dag}\kappa}_1 + \frac{\norm{\bar{\bm{\Phi}} \kappa - \gamma}_2^2}{2} \Bigg \rbrace,
\end{equation}
where the bracketed term is called the \textit{objective function}. To solve this
convex optimization problem we adopt a forward-backward splitting algorithm
\citep[\textit{e.g.}][]{[40]}. A full description of this algorithm applied in
the current context is outlined in \citet{[12]}.
\par
Let $f(\kappa) = \mu \norm{\bm{\Psi}^{\dag} \kappa}_1$ denote our prior term,
and $g(\kappa) = \norm{\bar{\bm{\Phi}} \kappa - \gamma}_2^2 / 2$ denote our data
fidelity term. Then our optimization problem can be re-written compactly as,
\begin{equation}
    \argminT_{\kappa} \big \lbrace f(\kappa) + g(\kappa) \big \rbrace.
\end{equation}
The forward-backward iteration step is then defined to be,
\begin{equation}
    \kappa^{(i+1)} = \text{prox}_{\lambda^{(i)}f} \bigg (\kappa^{(i)} - \lambda^{(i)} \nabla g(\kappa^{(i)}) \bigg ),
\end{equation}
for iteration $i$, with gradient,
\begin{equation}
    \nabla g(\kappa) = \bm{\bar{\Phi}}^{\dag} (\bm{\Phi} \kappa - \gamma).
\end{equation}
If the wavelet dictionary $\bm{\Psi}$ is a tight frame (i.e.
$\bm{\Psi}^{\dag}\bm{\Psi} = \mathbb{I}$) the proximity operator is given by,
\begin{equation}
    \text{prox}_{\lambda f}(z) = z + \bm{\Psi} \bigg ( \text{soft}_{\lambda\mu} (\bm{\Psi}^{\dag}z) - \bm{\Psi}^{\dag}z \bigg ),
\end{equation}
where $\text{soft}_{\lambda}(z)$ is the point-wise soft-thresholding operator
\citep{[40]} and $\lambda$ is a parameter related to the step-size (which is
in turn related to the Lipschitz differentiability of the log-prior) which should
be set according to \citet{[12]}. The iterative algorithm is given explicitly
in the primary iterations of algorithm \ref{alg:forward_back}. Adaptations for
frames which are not tight can be found in \citet{[12]} and are readily available
within our framework.
\par
Our algorithm has distinct similarities to the GLIMPSE algorithm presented
by \citet{[6]}, but crucially differs in several aspects. Most importantly
we formulate the problem in a hierarchical Bayesian framework which allows
us to recover principled statistical uncertainties. In addition to this we
include Bayesian inference of the regularization parameter, a robust estimate
of the noise-level (which can be folded into the hierarchical model), and we use
super-resolution operators instead of non-discrete fast Fourier transforms.


\begin{algorithm}
    \setstretch{1.2}
    \caption{Forward-backward analysis algorithm}
    \hspace*{\algorithmicindent} \textbf{Input:} $\gamma \in \mathbb{C}^M$,  $\kappa^{(0)} \in \mathbb{C}^N$,  $\lambda$,  $\mu^{(0)} = i = t = 0$, $T_1,T_2 \in \mathbb{R}_+$ \\
    \hspace*{\algorithmicindent} \textbf{Output:} $\kappa^{\text{map}} \in \mathbb{C}^N$, $\mu \in \mathbb{R}_{+}$ \\
    \textbf{Precomputation:}\\
    \textbf{Do:}
    \begin{algorithmic}[1]
        \State    Calculate $\kappa^{(t)} = \argminT_{\kappa}  \big \lbrace f(\kappa) +  g(\kappa) \big \rbrace$,
        \State    Update $\mu^{(t+1)} = \frac{(N/k)+\alpha-1}{f(\kappa^{(t)}) + \beta}$,
        \State    $t = t + 1$,
        \State    On convergence, $\mu$ becomes fixed.
    \end{algorithmic}
    \textbf{Until:} Iteration limit reached.\\
    \textbf{Primary Iterations:}\\
    \textbf{Do:}
    \begin{algorithmic}[1]
        \State   update $\nu^{(i+1)} = \kappa^{(i)} - \lambda \bm{\bar{\Phi}}^{\dagger}(\bm{\bar{\Phi}} \kappa^{(i)} - \gamma)$,
        \State   compute $\eta = \bm{\Psi}^{\dagger}\nu^{(i+1)}$,
        \State   update $\kappa^{(i+1)} = \nu^{(i+1)} + \bm{\Psi} ( \text{soft}_{\lambda\mu}(\eta)-\eta)$,
        \State   $i = i + 1$.
    \end{algorithmic}
    \textbf{Until:} Stopping criterion satisfied. \\
    \textit{i.e.} $\frac{\norm{\kappa^{(i)} - \kappa^{(i+1)}}_2}{\norm{\kappa^{(i)}}_2} < T_1$ and $\frac{\text{obj}(\kappa^{(i)}) - \text{obj}(\kappa^{(i+1)})}{ \text{obj}(\kappa^{(i)})} < T_2$.
    \label{alg:forward_back}
\end{algorithm}

\subsection{Reduced Shear}
Due to a degeneracy between $\gamma$ and $\kappa$ the true observable quantity
is in fact the \textit{reduced shear} $g$ \citep{[1]},
\begin{equation}
    g = \frac{\gamma}{1-\kappa}.
\end{equation}
Deep in the weak lensing regime one can safely approximate $\gamma
    \approx g \ll 1$ which ensures that  the optimization problem remains linear.
However, when reconstructing regions close to massive structures (galaxy
clusters) this approximation is no longer strictly valid and we must unravel
this additional factor. We adopt the procedure outlined in \citet{[3]}, which
we also outline schematically in Figure \ref{fig:schematic-of-reduced-shear}
\emdash this method can be found in detail in \citeauthor{[47]}, pg 153. We
find that these corrections typically converge after $\sim$ 5-10 iterations.

\begin{figure}
    \begin{center}
        \begin{tikzpicture}[node distance = 2cm, auto]
            \node [block, text width=8em] (1) {Let: $\gamma^{i=0} = g  $};
            \node [block, below of=1, node distance=1.6cm, text width=8em] (2) {Calculate: $\kappa^{(i)}$};
            \node [block, below of=2, node distance=1.6cm] (3) {Update: $\gamma^{i+1} = \gamma^{i} ( 1 - \kappa^{(i)})$};
            \node [decision, below of=3] (4) {Stopping Criterion Met?};
            \node [block, right of=2, node distance=3.0cm, text width=6em] (5) {$i \rightarrow i+1$};
            \node [block, below of=4] (4b) {Return: $\kappa^{(i)}$.};
            \path [line] (1) -- (2);
            \path [line] (2) -- (3);
            \path [line] (3) -- (4);
            \path [line,dashed] (4) -| node[right] {No}(5);
            \path [line,dashed] (5) -- (2);
            \path [line,dashed] (4) -- node[near start] {Yes}(4b);
        \end{tikzpicture}
        \caption{Schematic of reduced shear iterations. An initial guess of the MAP
            solution $\kappa^{\text{map}}_i$ is constructed, the current best shear estimates
            $\gamma_i$ are then used in tandem to construct a new estimate of the true shear
            field $\gamma_{i+1}$.}
        \label{fig:schematic-of-reduced-shear}
    \end{center}
\end{figure}

\subsection{Regularization Parameter Selection}
One key issue of sparsity-based reconstruction methods is the selection of the
regularization parameter $\mu$. Several methodologies have arisen \citep{[6],
    [9], [14], [15]} for selecting $\mu$, though often the regularization parameter
is chosen somewhat arbitrarily \emdash as the integrity of the MAP solution is
assumed to be weakly dependent on the choice of $\mu$. However, to extract
principled statistical uncertainties on the recovered images, one must select
this parameter in a principled statistical manner.
\par
We apply the hierarchical Bayesian formalism developed by \citet{[16]} \emdash
the details of which are elegantly presented by the authors. Though we will
outline roughly the underlying argument here.
\par
First define a sufficient statistic $f$ to be $k$-homogeneous if $\exists \; k
    \in \mathbb{R}_{+}$ such that,
\begin{equation}
    f(\eta x) = \eta^kf(x), \: \forall x \in \mathbb{R}^N, \: \forall \eta > 0.
\end{equation}
All norms, composite norms and composition of norms with linear operators are
1-homogeneous \emdash and so our $\ell_1$-norm has $k$ of 1. If a sufficient
statistic $f$ is $k$-homogeneous, then the normalization factor $C(\mu)$ of
$p(\kappa | \mu)$ is given by \citep{[16]},
\begin{equation} \label{eq:proposition}
    C(\mu) = A \mu^{-N/k},
\end{equation}
where $A$ is a constant independent from $\mu$. The proposed Bayesian inference
model then implements a gamma-type hyper-prior \emdash which is a typical hyper-prior
for scale-parameters,
\begin{equation} \label{eq:hyper-prior}
    p(\mu) = \frac{\beta^{\alpha}}{\Gamma(\alpha)}\mu^{\alpha - 1} e^{-\beta \mu} \mathbb{I}_{\mathbb{R}_+}(\mu),
\end{equation}
where without loss of generality $\alpha = \beta = 1$. The result is effectively
insensitive to their value (in numerical experiments values of $\alpha,
    \beta \in [10^{-2}, 10^{5}]$ produced essentially no difference in $\mu$).
\par
Now, let us extend the inference problem of the log-posterior to the case where
$\mu$ is an additionally unknown parameter. In this context we compute the joint
MAP estimator $(\kappa^{\text{map}}, \mu^{\text{map}}) \in \mathbb{C}^{N} \times
    \mathbb{R}_+$ which maximizes $p(\kappa, \mu | \gamma)$ such that,
\begin{equation}
    \mathbf{0}_{N+1} \in \partial_{\kappa, \mu} \log p(\kappa^{\text{map}}, \mu^{\text{map}} | \gamma),
\end{equation}
where $\mathbf{0}_{i}$ is the $i$-dimensional null vector and
$\partial_{s}h(s^{\prime})$ is the set of sub-gradients of function $h(s)$ at
$s^{\prime}$. This in turn implies both that,
\begin{equation} \label{eq:initial-opt}
    \mathbf{0}_N \in \partial_{\kappa} \log p(\kappa^{\text{map}}, \mu^{\text{map}} | \gamma),
\end{equation}
and
\begin{equation} \label{eq:new-opt}
    \mathbf{0} \in \partial_{\mu} \log p(\kappa^{\text{map}}, \mu^{\text{map}} | \gamma).
\end{equation}
\par
From equation (\ref{eq:initial-opt}) we recover the optimization problem with
known regularization parameter $\mu$ given in equation
(\ref{eq:optimization-problem}). However, from equations (\ref{eq:proposition},
\ref{eq:hyper-prior}, \ref{eq:new-opt}) it follows that the MAP regularization
parameter $\mu$ is given by \citep{[16]},
\begin{equation}
    \mu^{\text{map}} = \frac{\frac{N}{k} + \alpha - 1}{f(\kappa^{\text{map}}) + \beta},
\end{equation}
where we recall that $N$ is the total dimension of our convergence space.
\par
It is precisely this optimal $\mu$ value which we wish to use in our
hierarchical Bayesian model. Hereafter we drop the map superscript from $\mu$
for clarity. To calculate $\mu$ we perform preliminary iterations defined by:
\begin{equation}
    \kappa^{(t)} = \argminT_{\kappa}  \bigg \lbrace f(\kappa; \mu^{(t)}) + g(\kappa) \bigg \rbrace,
\end{equation}
where $g(\kappa)$ is our likelihood term and,
\begin{equation}
    \mu^{(t+1)} = \frac{\frac{N}{k}+\alpha-1}{f(\kappa^{(t)}) + \beta}.
\end{equation}
Typically we find that these preliminary iterations take $\sim$ 5-10 iterations
to converge, and recover close to optimal parameter selection for a range of test
cases -- note that here the optimal selection of $\mu$ is that which maximizes the
SNR of a recovered image.
\par
Another factor which can influence the quality of reconstructions is the selection
of wavelet dictionary. In this paper we consider Daubechies (8 levels) and SARA
dictionaries \citep{[39],[50]}, though a wide variety of wavelet
        dictionaries exist, see \textit{e.g.} starlets \citep{[17]}. The 8-level SARA
dictionary is a combination of the Dirac and Daubechies 1 to 8 wavelet dictionaries.
It is important to note that we use the SARA dictionary, not the complete SARA
scheme \citep{[39],[50]}, which involves an iterative re-weighting scheme that
is not considered here.

\subsection{Super-Resolution Image Recovery}
Gridding of weak lensing data is advantageous in that it can provide a good
understanding of the noise properties \emdash a necessary feature for principled
uncertainty quantification. However, an inherent drawback of projecting data
into a grid is the possibility of creating an incomplete space due to low
sampling density -- often referred to as masking. Decomposition of spin signals
on bounded manifolds is inherently degenerate \citep{bunn2003}; specifically the
orthogonality of eigenfunctions is locally lost at the manifold boundaries,
leading to signal leakage between Fourier (or on the sphere, harmonic) modes.
\par
One approach to mitigate this problem is to avoid the necessity of gridding
by substituting a \textit{non-uniform discrete Fourier transform} (NFFT) into the
RHS of equation (\ref{eq:measurement_operator}) as presented by \citet{[6]}.
A downside of this NFFT approach is that the noise is more difficult to handle,
leading to complications when considering uncertainty quantification. Another
approach is to perform super-resolution image recovery, which we present in the
context of our algorithm.
\par
Suppose the dimension of our gridded measurement space is $M$, as
before, and the desired dimension of our solution space is $N^{\prime}$, where
$N^{\prime} \geq N$. In this setting we have shear measurements $\gamma \in
    \mathbb{C}^{M}$ and recovered convergence $\kappa \in \mathbb{C}^{N^{\prime}}$.
Let us now define a \textit{super-resolution} (subscript SR) measurement operator
to be,
\begin{equation}
    \bm{\Phi}_{\text{SR}} = \bm{\mathsf{M}} \; \bm{\mathsf{F}}^{-1}_{\text{lr}} \; \bm{\mathsf{D}} \; \bm{\mathsf{Z}} \; \bm{\mathsf{F}}_{\text{hr}}
\end{equation}
where $\bm{\mathsf{F}}_{\text{hr}}$ is a high resolution (dimension $N^{\prime}$)
fast Fourier transform, $\bm{\mathsf{Z}} \in \mathbb{C}^{N \times N^{\prime}}$ is
a Fourier space down-sampling which maps $\tilde{\kappa}^{\prime} \in
    \mathbb{C}^{N^{\prime}}$ on to $\tilde{\kappa} \in \mathbb{C}^{N}$, where tilde
represents Fourier coefficients, $\bm{\mathsf{D}}$ is the planar forward model given
by equation (\ref{eq:forwardmodel}), and $\bm{\mathsf{M}}$ is a standard masking
operator. Finally, $\bm{\mathsf{F}}^{-1}_{\text{lr}}$ is a low resolution (dimension
$M$) inverse fast Fourier transform. For completeness the super-resolution adjoint
measurement operator is given by,
\begin{equation}
    \bm{\Phi}^{\dagger}_{\text{SR}} = \bm{\mathsf{F}}^{-1}_{\text{hr}} \; \bm{\mathsf{Z}}^{\dagger} \; \bm{\mathsf{D}}^{\dagger} \; \bm{\mathsf{F}}_{\text{lr}} \; \bm{\mathsf{M}}^{\dagger},
\end{equation}
where $\bm{\mathsf{M}}^{\dagger}$ is adjoint masking (gridding),
$\bm{\mathsf{D}}^{\dagger}$ is the adjoint of $\bm{\mathsf{D}}$ (which is
self-adjoint hence $\bm{\mathsf{D}}^{\dagger}=\bm{\mathsf{D}}$), and
$\bm{\mathsf{Z}}^{\dagger} \in \mathbb{C}^{M^{\prime} \times M}$ is
zero padding in Fourier space which acts by mapping $\tilde{\gamma} \in
    \mathbb{C}^{M}$ to $\tilde{\gamma}^{\prime} \in \mathbb{C}^{M^{\prime}}$. Note
that when considering the KS estimate in the super-resolution setting a rescaling
function to account for the different Fourier normalization factors must be
introduced (which we absorb into the Fourier operators).
As before, this super-resolution measurement operator is concatenated with the
inverse covariance weighting to
form an analogous composite operator $\bm{\bar{\Phi}}_{\text{SR}}$ which is used
throughout the following analysis.
\par
Conceptually super-resolution allows partial inpainting of higher resolution
Fourier modes. In this way one is able to recover high resolution structure for
images from comparatively low resolution datasets. Such high resolution structure
is of course dependent on the prior information injected when solving the inverse
problem. Interestingly this raises another consideration: in scenarios where the
pixel-level observation count is very low the noise level dilutes high frequency
components and can limit the efficacy of reconstruction algorithms. In such a setting
gridding observational data onto a lower resolution map, with inherently lower
pixel-level noise, and performing a super-resolution reconstruction can recover far
better estimates of the high frequency modes, and thus often recovers  greater
reconstruction fidelity.

\begin{figure}
    \begin{center}
        \begin{tikzpicture}[node distance = 2cm, auto]
            \node [block, text width=8em] (1) {Calculate MAP solution: $\kappa^{\text{map}}$};
            \node [block, below of=1, node distance=1.6cm, text width=8em] (2) {Construct surrogate: $\kappa^{\text{sgt}}$};
            \node [block, below of=2, node distance=1.6cm, text width=14em] (3a) {Calculate: $\mathcal{E} = f(\kappa^{\text{sgt}}) + g(\kappa^{\text{sgt}})$};
            \node [decision, below of=3a, node distance=1.6cm] (3) {$\mathcal{E} \leq \epsilon_{\alpha}^{\prime}$ ? };
            \node [block, below of=3, right of=3, text width=8em] (4) {$\mathcal{Z}$ is Physical};
            \node [block, below of=3, left of=3, text width=8em] (5) {$\mathcal{Z}$ is Inconclusive};

            \path [line] (1) -- node {Remove feature $\mathcal{Z}$} (2);
            \path [line] (2) -- (3a);
            \path [line] (3a) -- (3);
            \path [line,dashed] (3) -| node [near end] {No}(4);
            \path [line,dashed] (3) -| node [near end] {Yes}(5);

        \end{tikzpicture}
        \caption{Schematic of hypothesis testing. The feature $\mathcal{Z}$ is entirely
            general and can be constructed by any well defined operator on the MAP solution
            $\kappa^{\text{map}}$.}
        \label{fig:HypthosisTesting}
    \end{center}
\end{figure}

\section{Bayesian Uncertainty Quantification} \label{Bayesian Uncertainties}
Estimators recovered from algorithms of the form presented in the previous section
are MAP solutions to, in general, ill-conditioned inverse problems, and as such
have significant intrinsic uncertainty. Theoretically, MCMC techniques could be
applied to recover the complete posterior in the context of Gaussian \citep{[26],
    [43]} and sparsity-promoting \citep{[11],[44]} priors but these approaches are
computationally demanding for high dimensional problems where $N$ is large. As $N$
can easily be larger than $10^6$ (\textit{e.g.} when considering $1024 \times 1024$
resolution images), MCMC approaches are often not feasible.
\par
In \citet{[10]} a methodology based on probability concentration is presented, which
uses MAP estimators to estimate theoretically conservative approximate Bayesian
credible regions (specifically highest posterior density credible regions) of
the posterior, $p(\kappa|\gamma)$. As this approach requires only knowledge of the
MAP solution and the objective function, the Bayesian credible regions can be
approximated efficiently in high dimensional settings.

\subsection{Highest Posterior Density Regions }
A posterior credible region at confidence level $100(1-\alpha)\%$ is a sub-set
$C_{\alpha} \in \mathbb{C}^N$ which satisfies the integral,
\begin{equation} \label{CredibleIntegral}
    p(\kappa \in C_{\alpha}|\gamma) = \int_{\kappa \in \mathbb{C}^N} p(\kappa|\gamma)\mathbb{I}_{C_{\alpha}}d\kappa = 1 - \alpha,
\end{equation}
where $\mathbb{I}_{C_{\alpha}}$ is the set indicator function for $C_{\alpha}$
defined by $\mathbb{I}_{C_{\alpha}}(\kappa) = 1 \; \forall \kappa \in C_{\alpha}$
and $0$ elsewhere. One possible region which satisfies this property is the
\textit{Highest Posterior Density} (HPD) region defined by,
\begin{equation}
    C_{\alpha} := \lbrace \kappa : f(\kappa) + g(\kappa) \leq \epsilon_{\alpha} \rbrace,
\end{equation}
where $\epsilon_{\alpha}$ defines an iso-contour (\textit{i.e.} level-set) of the
log-posterior set such that the integral in (\ref{CredibleIntegral}) is satisfied.
This region can be shown \citep{[19]} to have minimum volume and is thus
decision-theoretically optimal. However, due to the dimensionality of the integral
in (\ref{CredibleIntegral}) calculation of the HPD credible region is difficult. A
conservative approximation of $C_{\alpha}$ was recently proposed \citep{[10]} and
shown to be effective in the inverse imaging setting of radio interferometric
imaging \citep{[12]}. This approximate HPD is defined by
\begin{equation}
    C^{\prime}_{\alpha} := \lbrace \kappa : f(\kappa) + g(\kappa) \leq \epsilon^{\prime}_{\alpha} \rbrace,
\end{equation}
where the approximate threshold $\epsilon^{\prime}_{\alpha}$ is given by
\begin{equation}
    \epsilon^{\prime}_{\alpha} = f(\kappa^{\text{map}}) + g(\kappa^{\text{map}}) + \tau_{\alpha} \sqrt{N} + N,
\end{equation}
with constant $\tau_{\alpha} = \sqrt{16 \log(3 / \alpha)}$. For a detailed
derivation of this approximation see \citet{[10]}. Provided $\alpha \in \big
    ( 4\exp(-N/3) \;, 1   \big )$ the deviation of this adapted threshold is bounded
and grows at most linearly with respect to $N$. The error of this approximate
threshold is bounded by
\begin{equation}
    0 \leq \epsilon^{\prime}_{\alpha} - \epsilon_{\alpha} \leq \eta_{\alpha} \sqrt{N} + N,
\end{equation}
where $\eta_{\alpha} = \sqrt{16 \log (3/\alpha)} + \sqrt{1/\alpha}$. In high
dimensional settings ($N$ large) this error may naively appear large, however
in practice the error is relatively small.

\subsection{Hypothesis Testing} \label{HypothesisTesting}
Extending the concept of HPD credible regions, one can perform \textit{knock-out}
hypothesis testing of the posterior to determine the physicality of recovered
structure \citep{[12]}.
\par
To perform such tests one first creates a surrogate image $\kappa^{\text{sgt}}$ by
masking a feature of interest $\Omega_D \subset \Omega$ in the MAP estimator
$\kappa^{\text{map}}$. It is then sufficient to check if,
\begin{equation}
    f(\kappa^{\text{sgt}}) + g(\kappa^{\text{sgt}}) \leq \epsilon_{\alpha}^{\prime}.
\end{equation}
If this inequality holds, we interpret that the physicality of $\Omega_D$ is
undetermined and so no strong statistical statement can be made. Should the
objective function evaluated at $\kappa^{\text{sgt}}$ be larger than
$\epsilon_{\alpha}^{\prime}$ then it no longer belongs to the approximate credible
set $C_{\alpha}^{\prime}$ and therefore (as $\epsilon_{\alpha}^{\prime}$ is
conservative) it \textbf{cannot} belong to the HPD credible set $C_{\alpha}$.
Therefore, for $\kappa^{\text{sgt}}$ which do not satisfy the above inequality
we determine the structure $\Omega_D$ to be strictly physical at $100(1-\alpha)\%$
confidence level. A schematic of hypothesis testing is provided in Figure
\ref{fig:HypthosisTesting}.
\par
In pixel-space we begin by masking out a feature of interest, creating a rough
surrogate image \emdash setting the pixels associated with a selected structure
to 0 \emdash this rough surrogate is then passed through an appropriate wavelet
filter $\Lambda$ as part of \textit{segmentation-inpainting} to replace generic
background structure into the masked region. Mathematically, this amounts to the
iterations,
\begin{equation}
    \kappa^{(i+1),\text{sgt}} = \kappa^{\text{map}}\mathbb{I}_{\Omega - \Omega_D} + \Lambda^{\dag}\text{soft}_{\lambda_t}(\Lambda \kappa^{(i),\text{sgt}})\mathbb{I}_{\Omega_D},
\end{equation}
where $\Omega_D \subset \Omega$ is the sub-set of masked pixels,
$\mathbb{I}_{\Omega-\Omega_D}$ is the set indicator function and $\lambda_t$ is a
thresholding parameter which should be chosen appropriately for the image.
\par
A second straightforward method for generating surrogate images is to blur local
pixel substructure into one collective structure \emdash in a process called
\textit{segmentation-smoothing}. This approach provides a simple way to determine
if the substructure in a given region is physical or likely to be an artifact of
the reconstruction process.
\par
For example, if several massive peaks are located near one another, one can blur
these structures into a single cohesive peak. This would be useful when considering
peak statistics on convergence maps \emdash which is often used to constrain the
cosmological parameters associated with dark matter.
\par
One can conduct such blurring of structure by: specifying a a subset of the
reconstructed pixels $\Omega_D \subset \Omega$; convolving $\kappa^{\text{map}}$
with a Gaussian smoothing kernel; and replacing pixels that belong to $\Omega_D$
with their smoothed counterparts. This can be displayed algorithmically as,
\begin{equation}
    \kappa^{\text{sgt}} = \kappa^{\text{map}}\mathbb{I}_{\Omega-\Omega_D} + \big (\kappa^{\text{map}} \ast \mathcal{G}(0,\chi) \big ) \mathbb{I}_{\Omega_D},
\end{equation}
where $\mathcal{G}(0,\chi)$ is a chosen Gaussian smoothing kernel and $\ast$ is a
trivially extended 2D version of the the usual 1D Fourier convolution operator,
\par
In the scope of this paper we focus primarily on pixel-space features, but it is
important to stress that \textit{knock-out} approach is entirely general and can
be applied to any well defined feature of a MAP estimator \emdash \textit{i.e.}
masking certain Fourier space features, removal of global small scale structure
\textit{etc.}

\section{Illustration on simulations} \label{Simulation Results}
We now consider a selection of realistic simulations to illustrate our
sparse reconstruction method on cluster scales which are particularly challenging for
myriad factors. Further to this, we showcase the aforementioned uncertainty
quantification methods in a variety of idealized cluster scale MAP reconstructions.
We place emphasis on
uncertainty quantification rather that the reconstruction fidelity.

\subsection{Datasets}
In this paper we focus primarily on 4 large clusters  (those with significant friends-of-friends,
\textit{i.e.} significant substructure) extracted from the Bolshoi N-body simulation \citep{[20]}.
On the cluster scale we showcase our formalism on a variety of Bolshoi N-body simulation data sets.
The Bolshoi N-body cluster simulation catalogs we work with in this paper are those used
in \citet{[6]}, which were extracted using the CosmoSim web-tool\footnote{https://www.cosmosim.org}.
Construction of these weak lensing realisations assumed a redshift of $0.3$, with a $10 \times 10$
arcmin$^2$ field of view, and have convergence normalized with respect to lensing
sources at infinity. Explicitly this results in pixel-dimensions of $\sim 2.5$ arcseconds.
Due to the relatively low particle density, these images were subsequently  denoised by a
multi-scale Poisson denoising algorithm.

\subsection{Methodology}
Typically, we begin by creating an artificial shear field $\hat{\gamma} \in
    \mathbb{C}^M$ from a known \textit{ground-truth} convergence field $\kappa$,
that is extracted from a given dataset. This is a common approach in the imaging
community and presents a closed scenario in which the true input is known. These
$\hat{\gamma}$ fields are created by,
\begin{equation}
    \hat{\gamma} = \bm{\Phi} \kappa + \mathcal{N}(0,\sigma_i^2),
\end{equation}
where $\sigma_i$ (\textit{i.e.} the noise covariance) is determined entirely from a pre-defined
number density of observations $n_{\text{gal}}$ per arcminute${}^2$, an assumed intrinsic
ellipticity dispersion of $0.37$, and the resolution of the images (in this case $10 \times 10$
arcminutes). In this way the noise can be tuned to directly mimic that present in practical settings.
Using the simulated noise covariance (which in practice would be provided by the observation
team) we then utilize the SOPT\footnote{A highly optimized sparse optimization solver,
    https://github.com/astro-informatics/SOPT} framework to perform our reconstruction algorithm
on $\hat{\gamma}$ such that we recover a MAP estimator of the convergence $\kappa^{\text{map}}$.
From this reconstructed convergence field a recovered SNR is computed and a selection
of hypothesis tests are conducted to showcase the power of this formalism.
\par
In the case where the underlying clean $\gamma$ are unavailable (\textit{i.e.}
application to A520 data) we conduct the same analysis as before but instead of
creating artificial noisy $\hat{\gamma}$ maps we used the real noisy observational
data.
\par
Throughout our analysis the recovered SNR (dB) is defined to be,
\begin{equation}
    \text{SNR} = 20 \times \log_{10}\Bigg (\frac{\norm{\kappa}_2}{\norm{\kappa - \kappa^{\text{map}}}_2} \Bigg ),
\end{equation}
when the ground-truth convergence is known. Furthermore we quantify the topological
similarity between the true convergence and the estimator \textit{via} the Pearson
correlation coefficient which is defined to be
\begin{equation}
    r = \frac{ \sum_{i=1}^{N_{\mathbb{S}^2}} \lbrace \kappa^{\text{map}}(i) - \bar{\kappa}^{\text{map}} \rbrace
    \lbrace \kappa(i) - \bar{\kappa} \rbrace }{ \sqrt{\sum_{i=1}^{N_{\mathbb{S}^2}} \lbrace \kappa^{\text{map}}(i) -
    \bar{\kappa}^{\text{map}} \rbrace^2} \sqrt{\sum_{i=1}^{N_{\mathbb{S}^2}} \lbrace \kappa(i) - \bar{\kappa} \rbrace^2}  },
\end{equation}
where $\bar{x} = \langle x \rangle$. The correlation coefficient $r \in [-1,1]$
quantifies the structural similarity between two datasets: 1 indicates maximal
positive correlation, 0 indicates no correlation, and -1 indicates maximal negative
correlation.

\begin{table}
    \centering
    \caption{Contains both reconstruction SNR and Pearson correlation coefficient (topological
        correlation) metrics for the raw KS (no smoothing), an optimally smoothed KS (grid search
        for smoothing kernel which maximizes the recovered SNR), and our sparse reconstructions
        of the Bolshoi-3 cluster simulated with realistic noise derived from the presented number density
        of galaxy observations $n_{\text{gal}}$. The difference column is calculated as the difference
        between the Sparse and smoothed KS recovered SNR. Note that dB is a logarithmic scale
        therefore increases of $\sim 20$dB are extreme reductions in RMS error.}
    \label{tab:SNR_data}
    \begin{tabular}{l|c|c|c|c|c|r} 
        \hline
        \hline
        \textbf{Input}   & \multirow{2}{*}{\textbf{KS}} & \textbf{KS}     & \multirow{2}{*}{\textbf{Sparse}} & \multirow{2}{*}{\textbf{Difference}} \\
        $n_{\text{gal}}$ &                              & \textbf{Smooth} &                                  &                                      \\
        \hline
        \hline
        \multicolumn{5}{|c|}{\textbf{SNR (dB)}}                                                                                                     \\
        \hline
        500              & 2.917                        & 6.276           & 27.506                           & + 21.230                             \\
        100              & -4.497                       & 5.774           & 21.955                           & + 16.181                             \\
        30               & -10.400                      & 5.340           & 21.462                           & + 16.122                             \\
        10               & -15.970                      & 5.041           & 14.409                           & + 9.368                              \\
        \hline
        \multicolumn{5}{|c|}{\textbf{Pearson Correlation}}                                                                                          \\
        \hline
        500              & 0.166                        & 0.902           & 0.977                            & + 0.075                              \\
        100              & 0.076                        & 0.796           & 0.970                            & + 0.174                              \\
        30               & 0.039                        & 0.689           & 0.955                            & + 0.266                              \\
        10               & 0.029                        & 0.716           & 0.949                            & + 0.233                              \\
        \hline
        \hline
    \end{tabular}
\end{table}
\begin{figure*}
    \centering
    \scalebox{0.97}{
    \includegraphics[width=0.6\columnwidth, trim={0.cm 2.0cm 18.8cm 2.1cm},clip]{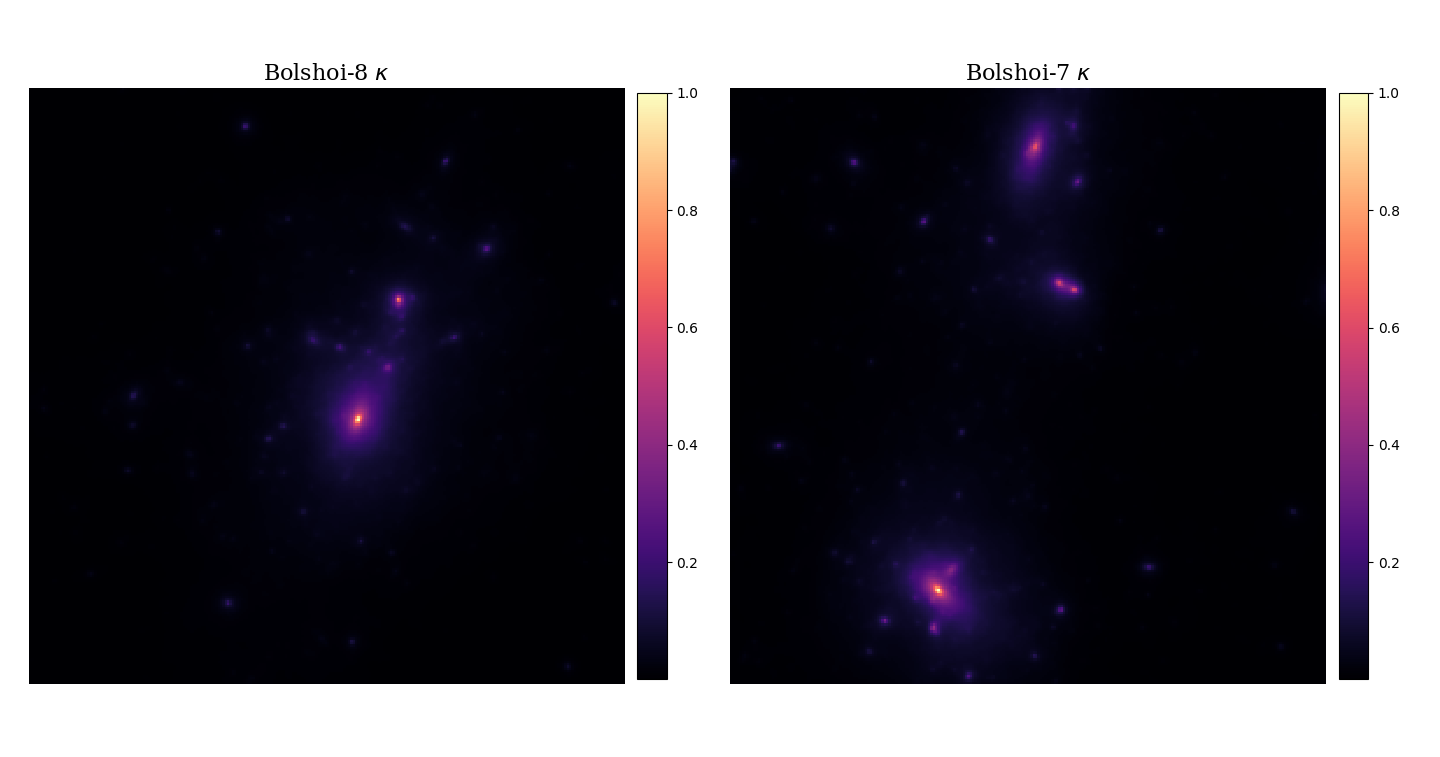}
    \put(-155,55){\textbf{\Large \rotatebox[origin=c]{90}{Ground Truth $\kappa$}}}}
    \scalebox{0.97}{
    \includegraphics[width=1.65\columnwidth, trim={0.0cm 2.1cm 1.65cm 1.5cm},clip]{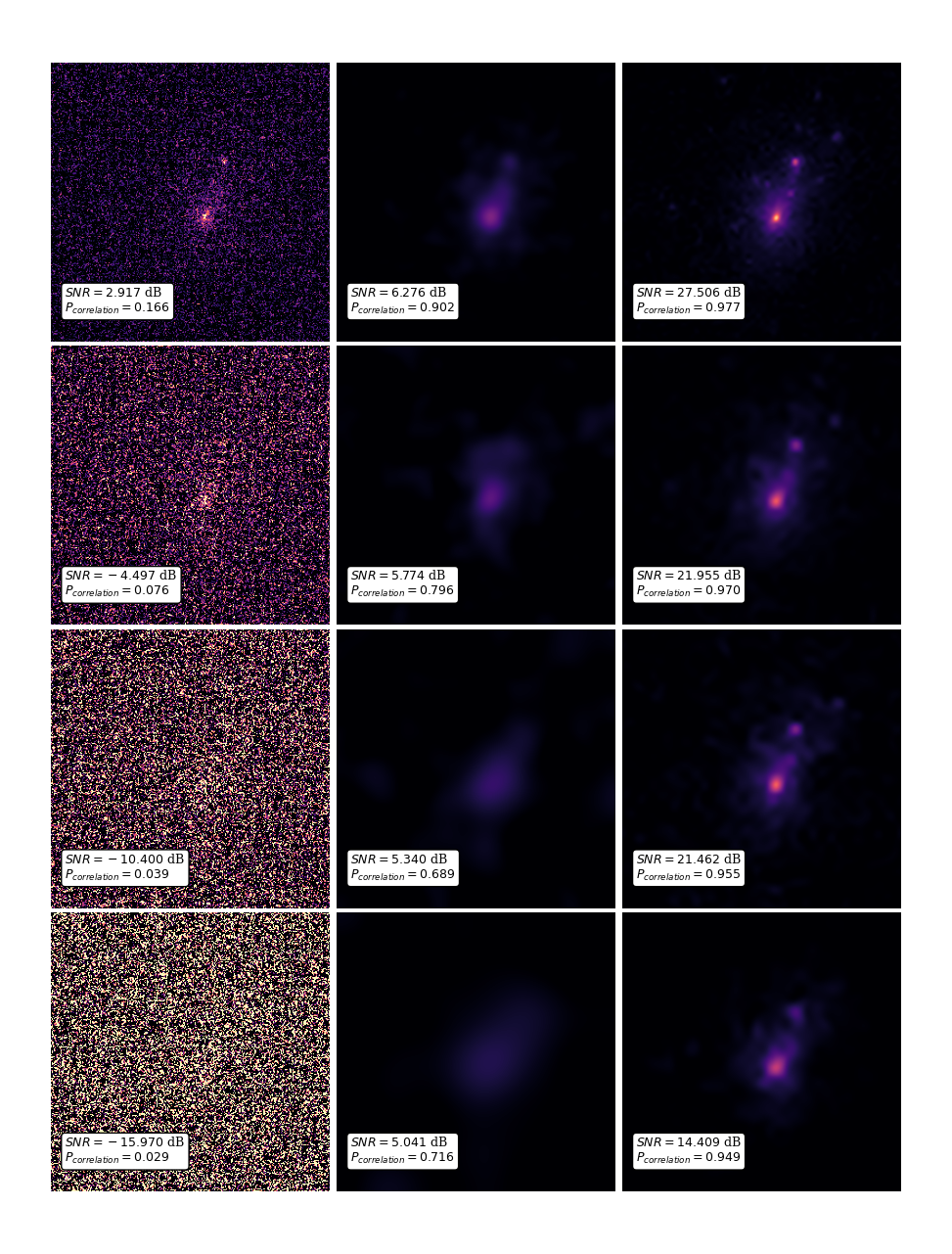}
    \put(-390,420){\textbf{\Large \rotatebox[origin=c]{90}{$n_{\text{gal}}$ = 500}}}
    \put(-390,300){\textbf{\Large \rotatebox[origin=c]{90}{$n_{\text{gal}}$ = 100}}}
    \put(-390,185){\textbf{\Large \rotatebox[origin=c]{90}{$n_{\text{gal}}$ = 30}}}
    \put(-390,55){\textbf{\Large \rotatebox[origin=c]{90}{$n_{\text{gal}}$ = 10}}}
    \put(-323,-10){\textbf{\Large KS}}
    \put(-213,-10){\textbf{\Large Smooth KS}}
    \put(-78,-10){\textbf{\Large Sparse}}}
    \newline
    \centering
    \caption{\textbf{Top to bottom:} Ground truth convergence map, simulations with noise levels
                corresponding to $n_{\text{gal}} \in [500, 100, 30, 10]$ respectively. Notice the clear effectiveness
                of sparse reconstruction over the standard KS method for a range of input SNR values.
                The numerical details can be found in Table \ref{tab:SNR_data}. The vertical labels indicate
                the input $n_{\text{gal}}$ used to simulate realistic noise for a given row, whereas
                horizontal labels indicate the reconstruction type. An optimal (grid searched
                to maximize the recovered SNR) Gaussian smoothing kernel was applied to the KS recovery
                to yield the KS (smooth) recovery in an attempt to remove noise from the KS
                estimator (obviously this is not possible in practice, where the ground truth is
                unknown: results shown therefore present the best possible performance for the
                smoothed KS estimator). Clearly, in all cases, the super-resolution sparse
                approach produces convergence maps which are far more representative of the ground truth
                across the aforementioned metrics.}
    \label{fig:bolshoi_3_SNR_figure_no_super_res}
\end{figure*}

\begin{figure*}
    \centering
    \includegraphics[width=\textwidth, trim={0.2cm 0.1cm 2.0cm 0.7cm},clip]{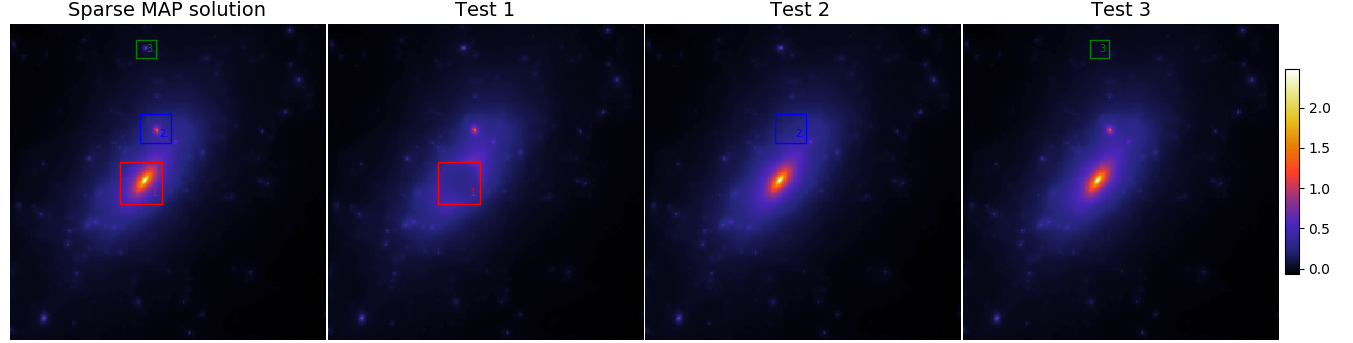}
    \put(-448,130){\textbf{\Large $\bm{\kappa}^{\text{map}}$}}
    \put(-325,130){\textbf{\Large H1}}
    \put(-200,130){\textbf{\Large H2}}
    \put(-75,130){\textbf{\Large H3}}

    \includegraphics[width=\textwidth, trim={0 0.2cm 2.7cm 0.7cm},clip]{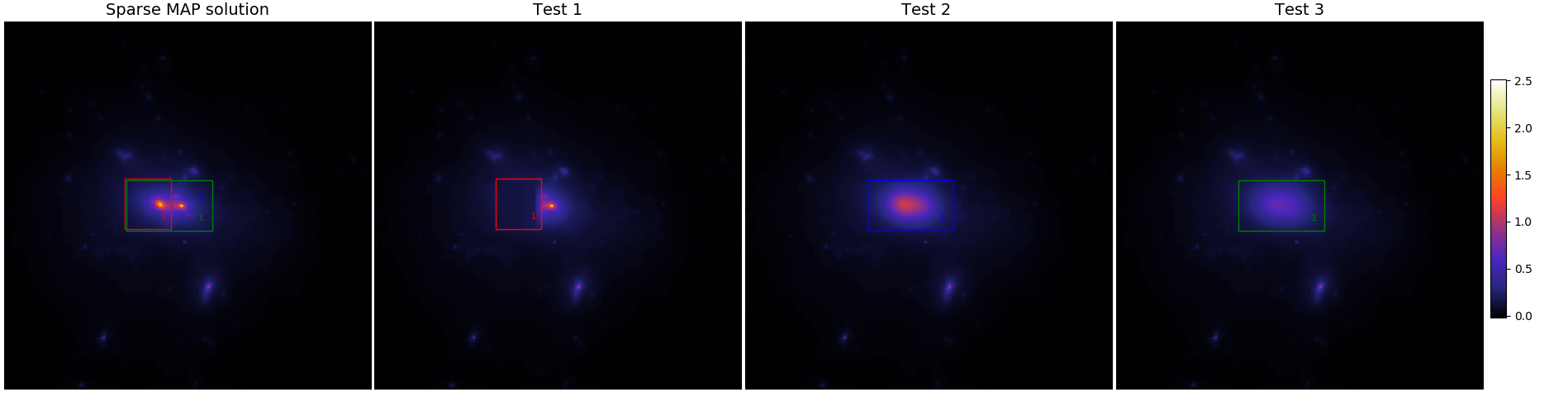}

    \includegraphics[width=\textwidth, trim={0.95cm 0.2cm 3.5cm 1.8cm},clip]{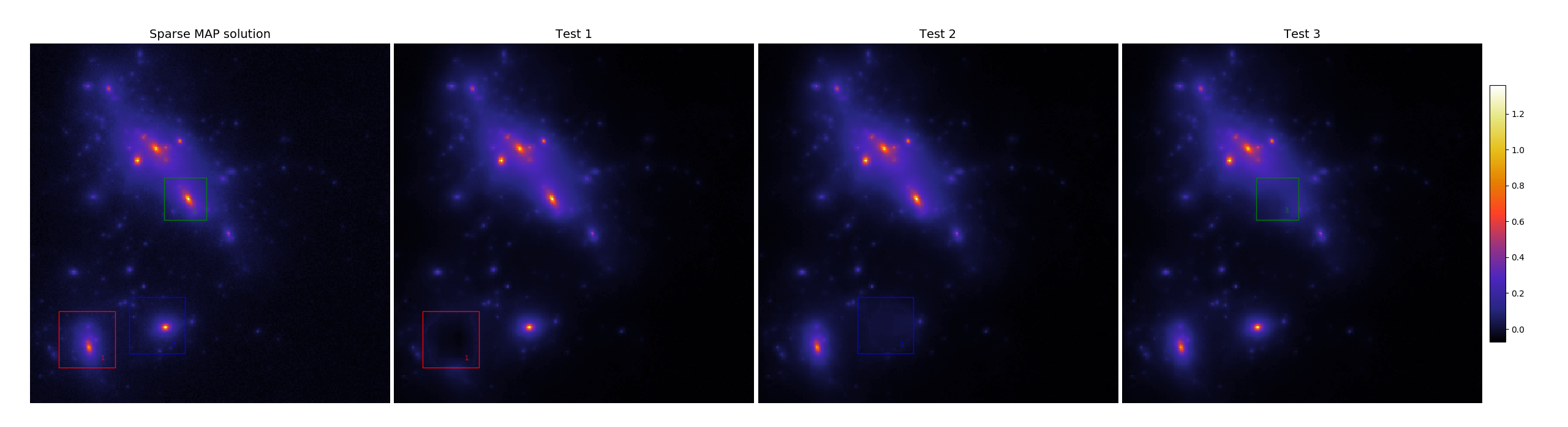}
    \caption{\textbf{General:} Hypothesis testing of three selected structures in the Bolshoi-1 cluster convergence
                field. The number density of galaxy observations $n_{\text{gal}}$ was set idealistically to 500 arcminute$^{-2}$
                simply for demonstration purposes. Additionally super-resolution
                was not active and the masking was trivially set to the identity, again to simplify the example
                for demonstration purposes. All numerical details can be found in Table \ref{tab:bolshoi_hyp_data}.
                \textbf{Top row:} We correctly determine that region 1 (\textit{red}) is physical with $99\%$ confidence. Regions 2 (\textit{blue})
                and 3 (\textit{green}) remain within the HPD region and are therefore inconclusive, given the data and
                noise level.
                \textbf{Middle row:} We correctly determine that all three null hypotheses (\textit{red, blue} and \textit{green}) are
                rejected at $99\%$ confidence. In H1 the conclusion is that the left hand peak was statistically
                significant. In H2 and H3 the conclusions is that an image with the two peaks merged it
                unacceptable, and therefore the peaks are distinct at $99\%$ confidence.
                \textbf{Bottom row:}  We correctly determine that all three hypothesis regions (\textit{red, blue} and \textit{green})
                $\Omega_D$ are physical with $99\%$ confidence.
                }
    \label{fig:bolshoi_1_hypothesis_testing}
\end{figure*}

\subsection{Bolshoi Cluster Catalogs} \label{bolshoi}
The Bolshoi cluster data used consists of 4 large clusters extracted
from the Bolshoi N-body simulation \citep{[20],[6]}. These images were then
multi-scale Poisson denoised to create suitable ground truth simulations. We
choose to analyze the same clusters considered in \citet{[6]}, as they showcase
a wide variety of structure on all scales. Hereafter, we restrict ourselves to
the SARA dictionary \citep{[39]} truncated at the $4^{\text{th}}$ Daubechies
wavelet (DB4) for simplicity \emdash \textit{i.e.} the combination of the Dirac,
and DB1 to DB4 wavelet dictionaries only.

To investigate the SNR gain of our formalism over KS in the cluster scale setting,
we created realizations of noisy pseudo-shear maps for assumed number density of
galaxy observations $n_{\text{gal}} \in [500,100,30,10]$ from one Bolshoi cluster
map, upon which we applied our reconstruction algorithm pipeline. The results of
which are presented in Table \ref{tab:SNR_data}. It should be noted that for
comparisons sake the KS estimate without convolution with a Gaussian smoothing
kernel is provided in addition to an optimally smoothed KS estimator. This has
been done to highlight the difference in reconstruction fidelity between the raw
KS estimator and the KS estimator after post-processing (Gaussian smoothing), a
discrepancy often not addressed by the community. As this post-processing
convolution is known to degrade the quality of non-Gaussian information (which
cosmologists are becoming increasingly interested in) such plots demonstrate
the trade-off between non-Gaussian information and reconstruction fidelity.

\par
As can be seen in Figure \ref{fig:bolshoi_3_SNR_figure_no_super_res} and Table
\ref{tab:SNR_data}, sparse approaches significantly outperform the smoothed (and non-smoothed) KS approach
in all cases, over all metrics tracked. Importantly sparse approaches are able to recover
reasonable results even when the noise level entirely dilutes the true signal, as in the
$n_{\text{gal}} = 10$ setting, making such approaches on (at least) cluster data very
attractive for future studies.

\subsubsection{Hypothesis Testing: Bolshoi Clusters}
Perhaps more interestingly, we now perform a series of hypothesis tests as discussed
in Section \ref{HypothesisTesting}. For each of the remaining 3 Bolshoi cluster we
construct three possible example hypothesis tests which one may wish to perform.
In this case these hypotheses were either: structure removal followed by
segmentation-inpainting; or Gaussian smoothing of certain structures (\textit{i.e.}
smoothing multiple peaks into a single larger peak which may be of interest when
conducting peak-count analysis). Though these are both extremely useful
considerations, it is important to stress the generality of our approach such
that any well defined operation on the reconstructed image, with a clear
understandable hypothesis, is applicable.
\par
To ensure the method behind hypothesis testing is clear, we will walk through a
typical application. The top row of Figure \ref{fig:bolshoi_1_hypothesis_testing}
displays the hypothesis tests applied to the first Bolshoi cluster. Conceptually,
the correct way to interpret Hypothesis 1 (H1, red) is: \textit{`The central dark
    core is likely just an artifact of the reconstruction'}.
\par
This structure is then removed from the image by segmentation-inpainting (lower
left image), and the objective function is then recalculated. It is found that
the objective function is now larger than the approximate level-set threshold
$\epsilon^{\prime}_{99\%}$, the surrogate segmentation-inpainted image falls
outside of the $99\%$ HPD credible region, and so the hypothesis is
\textbf{rejected}. This implies that the structure is not simply an artifact,
but is necessary to the integrity of the reconstruction, \textit{i.e.} this
structure is now determined to be physical at $99\%$ confidence. However, had
removing this region \textbf{not} raised the objective function above
$\epsilon^{\prime}_{99\%}$, then the conclusion is that their is insufficient
evidence to reject the hypothesis (which is \textbf{not} equivalent to saying
that the region is strictly not physical).
\par
An identical thought process can be applied to H2 and H3 of the top row in
Figure \ref{fig:bolshoi_1_hypothesis_testing}, H1 in the second row of Figure
\ref{fig:bolshoi_1_hypothesis_testing}, and all three hypothesis tests presented
in the final row. In each case a substructure of the $\kappa^{\text{map}}$ is
removed \emph{via} segmentation-inpainting and it is queried whether the resulting
surrogate solution $\kappa^{\text{sgt}} \in C_{\alpha}^{\prime}$. Each of the
large substructures on the final row, and H2 of the second row, are determined
to be physical at $99\%$ confidence. Conversely, the comparatively smaller
substructures considered in H2 and H3 of the top row do not saturate the level-set
threshold, and are therefore undetermined. All numerical data related to hypothesis
testing of the Bolshoi cluster reconstructions can be found in Table
\ref{tab:bolshoi_hyp_data}.
\par
H2 and H3 of the middle row of Figure \ref{fig:bolshoi_1_hypothesis_testing}
have a different interpretation. In these cases the central region has been
blurred by segmentation-smoothing (convolution with a Gaussian smoothing kernel)
\emdash the difference between these two cases being simply the degree of
smoothing. Here the hypothesis is: \textit{`The central region is likely to be
    just a single peak, rather than two'}. As in the previous example, the objective
function is recalculated and is now greater than $\epsilon^{\prime}_{99\%}$ and
so the hypothesis is rejected. The natural conclusion is thus that the data
\textbf{is} sufficient to determine that at least two peaks are physically
present at $99\%$ confidence.

\begin{table}
    \caption{Displays the MAP objective function, level-set threshold at 99\% confidence,
        surrogate objective function and whether the removed region was successfully
        identified as being physical. This data-set corresponds to Figures\ref{fig:bolshoi_1_hypothesis_testing}}
    \label{tab:bolshoi_hyp_data}
    \begin{tabular}{lccccr} 
        \hline
        \hline
        \multirow{2}{*}{\textbf{Test}} & \textbf{Initial}        & \textbf{Threshold}         & \textbf{Surrogate}                            & \textbf{Reject}       \\
                                       & $f(\kappa) + g(\kappa)$ & $\epsilon_{99\%}^{\prime}$ & $f(\kappa^\text{sgt}) + g(\kappa^\text{sgt})$ & $\bm{H_0}$ \textbf{?} \\
        \hline
        \hline
        \multicolumn{5}{|c|}{\textbf{Bolshoi-1}}                                                                                                                      \\
        \hline
        H1                             & 95426                   & 163408                     & 805513                                        & \checkmark            \\
        H2                             & 95426                   & 163408                     & 134080                                        & $\times$              \\
        H3                             & 95426                   & 163408                     & 100582                                        & $\times$              \\
        \hline
        \multicolumn{5}{|c|}{\textbf{Bolshoi-2}}                                                                                                                      \\
        \hline
        H1                             & 97121                   & 165103                     & 824260                                        & \checkmark            \\
        H2                             & 97121                   & 165103                     & 221492                                        & \checkmark            \\
        H3                             & 97121                   & 165103                     & 366981                                        & \checkmark            \\
        \hline
        \multicolumn{5}{|c|}{\textbf{Bolshoi-3}}                                                                                                                      \\
        \hline
        H1                             & 83419                   & 151401                     & 369939                                        & \checkmark            \\
        H2                             & 83419                   & 151401                     & 234305                                        & \checkmark            \\
        H3                             & 83419                   & 151401                     & 314089                                        & \checkmark            \\
        \hline
        \hline
    \end{tabular}
\end{table}
\begin{table}
    \caption{Displays the MAP objective function, level-set threshold at 99\% confidence, surrogate
                objective function and whether the null hypothesis $H_0$ is rejected. As can be seen, both MAP
                solutions fail to reject the null hypothesis in the other's objective function. This leads us to
                conclude that the two datasets do not disagree at 99\% confidence. Further discussion akin
                to the Kullback-Leibler divergence of the two posteriors is beyond the scope of this paper, but
                perhaps of interest in future work.}
    \label{tab:C12_J14_hypothesis_data}
    \begin{tabular}{cccccc} 
        \hline
        \hline
        \textbf{Hypothesis}                                   & \textbf{Initial}      & \textbf{Threshold}         & \textbf{Surrogate}                          & \textbf{Reject}      \\
        \textbf{Test}                                         & $f(\kappa)+g(\kappa)$ & $\epsilon_{99\%}^{\prime}$ & $f(\kappa^\text{sgt})+g(\kappa^\text{sgt})$ & $\bm{H_0}$\textbf{?} \\
        \hline
        \hline
        \citetalias{[36]} $\Leftrightarrow$ \citetalias{[32]} & 99231                 & 168044                     & 125601                                      & $\times$             \\
        \citetalias{[32]} $\Leftrightarrow$ \citetalias{[36]} & 98943                 & 167243                     & 134391                                      & $\times$             \\
        \hline
    \end{tabular}
\end{table}

\section{Application to Abel-520 Observational Catalogs} \label{A520}
We perform an application of our entire reconstruction pipeline to real
observational datasets. We select two observational datasets of the A520
cluster \citep{[32],[36]} \emdash hereafter for clarity we refer to them as
\citetalias{[36]} and \citetalias{[32]}
\citep[as in ][]{[9]}\footnote{http://www.cosmostat.org/software/glimpse}.
For a full description of the datasets, how they were constructed, and how
they account for different systematics we recommend the reader look to the
respective papers. These initial investigations claim to have
detected several over dense regions within the merging A520 system, the most
peculiar of which was a so called `dark core' (location 2 in Figure
\ref{fig:J14_C12_reconstructions}) for which multi-wavelength observations could
not determine an optical counterpart. Such a dark core would provide a
contradiction to the currently understood model of collisionless dark matter --
the idea being that during the collision of two massive clusters, dark matter
was stripped from each cluster through self-interactions, forming an over dense
residual between the two clusters, which would naturally not exhibit an optical
counterpart.
\par
The \citetalias{[32]} catalog contains approximately twice the number of
galaxies than \citetalias{[36]}, though both are derived from the same ACS
(four pointings)  and Magellan images. In addition, \citetalias{[32]} combines
these images with the CFHT catalog used in the authors previous work \citep{[37]}.
The \citetalias{[36]} observing area extends over a larger angular surface than
the \citetalias{[32]} so for this analysis we limit both datasets to the region
spanned by both sets. Due to the number density of measurements being very low we
are forced to project the measurements into a $32\times32$ grid \emdash to ensures
that the average number of galaxies in each grid pixel is at least above 1,
though ideally we want many galaxies in each pixel to minimize the noise
contribution from intrinsic ellipticity. In fact, even in this resolution the
space is incomplete in several pixels, but we draw a compromise between the
completeness of the space and the resolution of the data.
\par
The data covariance was constructed directly from the number density of
observations per pixel (directly inferred during catalogue gridding), with an assumed intrinsic ellipticity
dispersion of $0.37$. Combining this data covariance, the associated gridded datasets, and the
associated mask, MAP reconstructions of the \citetalias{[36]} and \citetalias{[32]} convergence
maps were recovered at a super-resolution magnification of 8. Reconstructions are
presented in Figure \ref{fig:J14_C12_reconstructions}.

\subsection{Hypothesis Testing of Local Structure: A520 Datasets}
\begin{figure}
    \centering
    \includegraphics[width=\columnwidth, trim={10cm 0cm 10cm 0cm},clip]{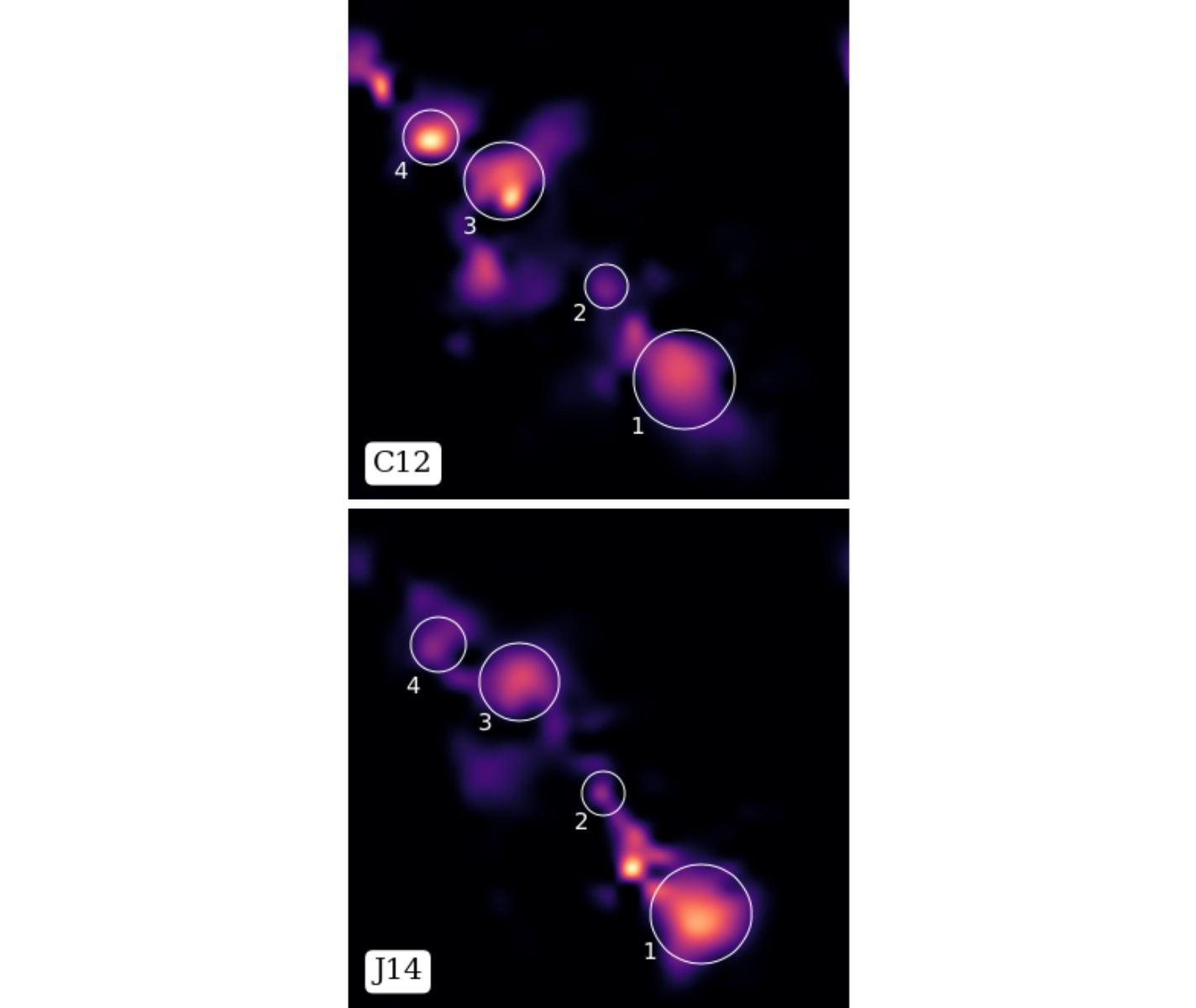} \\
    \caption{\textbf{Top:} Super-resolution sparse Bayesian
                reconstruction of \citetalias{[32]} and \citetalias{[36]} respectively. In a
                Bayesian manner it is found that the two datasets do not globally disagree
                at $99\%$ confidence. However, given the data resolution and noise-levels,
                only peaks 1 and 3 (in both datasets) could be determined to be statistically
                significant. This is \textbf{not} to say they do not exist, but implies that
                the data quantity and quality is insufficient to make a robust, principled
                statistical statement which could be used as evidence of their existence. The
                contested peak 2 is not detected at any reasonable confidence in either dataset.}
    \label{fig:J14_C12_reconstructions}
\end{figure}

We conducted hypothesis tests on the three primary over dense regions,
in addition to the contested dark core, in both the \citetalias{[36]}
and \citetalias{[32]} datasets. In the absence of an optical
        counterpart, detection at high confidence of the dark core (location 2 in
        Figure \ref{fig:J14_C12_reconstructions}) would provide a contradiction to
        the collisionless model of dark matter -- indicating potential self-interaction of
        dark matter. Due to the high estimated noise-level present in the data, and the
limited data resolution, only the two largest peaks in both datasets (peaks 1 and
3 of Figure \ref{fig:J14_C12_reconstructions}) sufficiently raised the objective
function to reject the hypothesis at any meaningful confidence. This is to say
that; given the limited, noisy data and using the measurement operator and prior
($\ell_1$-term) presented in this paper we can say that the data is insufficient
to statistically determine the physicality of local small scale substructure 
(such as the dark core) in both the \citetalias{[36]} and
\citetalias{[32]} datasets. The initial conflict between \citetalias{[36]}
        and \citetalias{[32]} was over the existence and position of a dark core (location
        2 in Figure \ref{fig:J14_C12_reconstructions}), with a notably large mass-to-light
        ratio, indicated the possibility of self-interacting dark matter. A subsequent
        inquiry was conducted \citep{[9]} using the GLIMPSE reconstruction algorithm
        \citep{[6]} and concluded that this peculiar peak existed in the \citetalias{[32]}
        dataset but not in the \citetalias{[36]} dataset.
\par
As such, our conclusions agree well with \citeauthor{[9]} (and generally with
those drawn in both \citetalias{[36]} and \citetalias{[32]}). However, within
our Bayesian hierarchical formalism (which constitutes a principled statistical
framework) we push this conclusion further to say that the data are insufficient
to determine the physicality of these peaks.

\subsection{Hypothesis Testing of Global Structure: A520 Datasets}

Interestingly we can perform a final novel hypothesis test of global structure. This hypothesis
is as follows: \textit{`The two MAP estimates are consistent with both sets of data',} \textit{i.e.} the
MAP convergence estimate recovered from the \citetalias{[32]} (\citetalias{[36]}) data is
within the credible-set (at $99\%$ confidence) of the \citetalias{[36]} (\citetalias{[32]}) objective
function. We find that the \citetalias{[32]} (\citetalias{[36]}) MAP reconstruction is an acceptable
solution to the \citetalias{[36]} (\citetalias{[32]}) inverse problem and so the MAP solutions do
not disagree \emdash numerically this is shown in Table \ref{tab:C12_J14_hypothesis_data}.

\par
Given the inherent limitations of the data we are forced to conclude: \textit{`The data are
    insufficient to determine the existence of individual substructures at high confidence \emdash though
    the two largest over dense regions are found to be globally physical at $99\%$ confidence.
    The two MAP estimates are also found to be consistent at $99\%$ confidence.'}

\section{Conclusions} \label{Conclusions}
We have presented a sparse hierarchical Bayesian mass-mapping algorithm which provides a principled statistical
framework through which, for the first time, we can conduct uncertainty quantification on recovered
convergence maps without relying on any assumptions of Gaussianity. Moreover, the presented
formalism draws on ideas from convex optimization (rather than MCMC techniques) which makes
it notably fast and allows it to scale well to big data, \textit{i.e.} high resolution and wide-field
convergence reconstructions (which will be essential for future stage \rom{4} surveys, such
as LSST and Euclid).
\par
Additionally, we demonstrate a hierarchical Bayesian inference approach to automatically
approximate the regularization parameter, and show that it produces near optimal results in
a variety of cases. We however note that this approach does not work generally, and can
be unstable in extreme settings.
\par
We showcase our Bayesian inference approach (with emphasis on the application of the
uncertainty quantification techniques) on both simulation datasets and observational data (the A520 merging cluster dataset).
Our mass-mapping formalism is shown to produce significantly more accurate convergence
reconstruction than the Kaiser-Squires estimator on all simulations considered. Hypothesis tests of
substructure are demonstrated.
\par
It is found that neither of the two A520 datasets considered could provide sufficient evidence
to determine the physicality of any contested substructure
(\emph{i.e.} the existence of so called `dark cores') at significant confidence. 
It is informative to note that our methods were, in fact, sufficiently
sensitive to detect the largest peaks in both datasets at $99\%$ confidence.
Nonetheless, global hypothesis tests indicate a good agreement between the two sets
of data. These conclusions are roughly in agreement with those drawn previously but
go further to demonstrate just how uncertain these types of cluster-scale weak lensing
reconstruction inherently are (typically as a limitation of the relative information
content of low-resolution, noisy datasets).
\par
It is now natural to extend this formalism to the entire celestial sphere \emdash a necessity
of large-scale reconstruction techniques which aim to fully utilize the forthcoming Euclid and
LSST\footnote{https://www.lsst.org} survey data.

\section*{Acknowledgements}

Author contributions are summarized as follows.
MAP: methodology, data curation, investigation, software, visualization, writing - original draft;
JDM: conceptualization, methodology, project administration, supervision, writing - review \& editing;
XC: methodology, investigation, writing - review \& editing;
TDK: methodology, supervision, writing - review \& editing;
CGRW: methodology.

This paper has undergone internal review in the LSST Dark Energy Science Collaboration. The internal
reviewers were Chihway  Chang, Tim Eifler, and François Lanusse. The authors would like to thank
Luke Pratley for an introduction to the C++ SOPT framework and the internal reviewers for valuable
discussions. MAP is supported by the Science and Technology Facilities Council (STFC).  TDK is
supported by a Royal Society University Research Fellowship (URF).  This work was also supported
by the Engineering and Physical Sciences Research Council (EPSRC) through grant EP/M0110891 and
by the Leverhulme Trust. The DESC acknowledges ongoing support from the Institut National de
Physique Nucl\'eaire et de Physique des Particules in France; the Science \& Technology Facilities
Council in the United Kingdom; and the Department of Energy, the National Science Foundation, and
the LSST Corporation in the United States.  DESC uses resources of the IN2P3 Computing Center
(CC-IN2P3--Lyon/Villeurbanne - France) funded by the Centre National de la Recherche Scientifique;
the National Energy Research Scientific Computing Center, a DOE Office of Science User Facility
supported by the Office of Science of the U.S.\ Department of Energy under Contract No.\ DE-AC02-05CH11231;
STFC DiRAC HPC Facilities, funded by UK BIS National E-infrastructure capital grants; and the UK
particle physics grid, supported by the GridPP Collaboration.  This work was performed in part
under DOE Contract DE-AC02-76SF00515.



\input{arxiv_resubmission_2021_15_5_MAP.bbl}

\bsp	
\label{lastpage}
\end{document}